\documentclass[letter,11pt]{article}
\usepackage{jheppub}
\usepackage{tikz}
\usetikzlibrary{decorations.pathmorphing}
\usetikzlibrary{decorations.markings}
\usepackage{xspace}
\usepackage{amsmath}
\usepackage{amssymb}
\usepackage{mathtools}
\usepackage[utf8]{inputenc}
\usepackage{physics}
\usepackage{slashed}
\usepackage{hyperref}
\usepackage[capitalize]{cleveref}

\newcommand{\API}{{\tt API}}
\newcommand{\JSON}{{\tt JSON}}
\newcommand{\hightea}{{\tt HighTEA}}

\title{\hightea : High energy Theory Event Analyser}

\author[a]{Micha\l{} Czakon,}
\author[b]{Zahari Kassabov,}
\author[c]{Alexander Mitov,}
\author[c]{Rene Poncelet,}
\author[c]{Andrei Popescu}

\affiliation[a]{Institut f\"ur Theoretische Teilchenphysik und Kosmologie, RWTH Aachen University, D-52056 Aachen, Germany}
\affiliation[b]{DAMTP, University of Cambridge, Wilberforce Road, Cambridge, CB3 0WA, United Kingdom}
\affiliation[c]{Cavendish Laboratory, University of Cambridge, Cambridge CB3 0HE, United Kingdom}

\emailAdd{mczakon@physik.rwth-aachen.de}
\emailAdd{zk261@cam.ac.uk}
\emailAdd{adm74@cam.ac.uk}
\emailAdd{poncelet@hep.phy.cam.ac.uk}
\emailAdd{andrei.popescu@cantab.net}

\preprint{Cavendish-HEP-23/01, P3H-23-021, TTK-23-08}

\abstract{We introduce \hightea, a new paradigm for deploying fully-differential next-to-next-to leading order (NNLO) calculations for collider observables. In principle, any infrared safe observable can be computed and, with very few restrictions, the user has complete freedom in defining their calculation's setup. For example, one can compute generic $n$-dimensional distributions, can define kinematic variables and factorization/renormalization scales, and can modify the strong coupling and parton distributions. \hightea\ operates on the principle of analyzing precomputed events. It has all the required hardware and software infrastructure such that users only need to request their calculation via the internet before receiving the results, typically within minutes, in the form of a histogram. No specialized knowledge or computing infrastructure is required to fully utilize \hightea, which could be used by both experts in particle physics and the general public. The current focus is on all classes of LHC processes. Extensions beyond NNLO, or to $e^+e^-$ colliders, are natural next steps.}

\begin{document} 
\maketitle
\flushbottom

\section{Introduction}\label{sec:intro}

The so-called {\it NLO revolution}, which peaked about 10 years ago, resulted in two important developments: first, the technology for performing any next-to-leading (NLO) calculation was established and refined and, second, a number of flexible, publicly available computer codes \cite{Frixione:1995ms,Catani:1996vz, Nagy:2003tz,Campbell:2011bn,Campbell:2015qma,Campbell:2019dru} and general purpose event generators \cite{Frixione:2007vw,Alwall:2014hca,Sjostrand:2014zea,Gleisberg:2008ta,Sherpa:2019gpd} appeared. The constant flow of ever-increasing-in-precision LHC measurements, however, quickly made it clear that the demand was shifting towards calculations with precision higher than NLO. 

Calculations with next-to-next-to-leading (NNLO) accuracy satisfy this requirement for almost all LHC processes, while for selected few cases N$^3$LO accuracy is required and has been achieved \cite{Anastasiou:2015vya,Chen:2021isd,Camarda:2021ict,Chen:2021vtu,Chen:2022cgv,Neumann:2022lft,Duhr:2020sdp,Baglio:2022wzu,Camarda:2023dqn}. The technology for performing NNLO calculations for a generic LHC process has already been developed and is fairly well established \cite{Gehrmann-DeRidder:2005btv,Catani:2007vq,Czakon:2010td,Czakon:2014oma,Gaunt:2015pea,Boughezal:2015dva,Currie:2016bfm,DelDuca:2016csb,Caola:2017dug,Magnea:2018hab}.  It has resulted in the calculation of almost all $2\to 2$ and many $2\to3$ processes. What is lacking at present is the possibility for easily obtaining such results. To be specific, NNLO calculations at present are being performed either with non-public codes (which are not directly accessible by general users), or with a small number of public codes \cite{Boughezal:2016wmq,Grazzini:2017mhc} which can compute an important but nonetheless limited range of processes. Running such programs already requires a relatively high level of sophistication on the side of the user. Additionally, NNLO calculations are computationally very expensive and the vast majority of users do not have such computational resources at their disposal. For example, a reasonably straightforward NNLO calculation takes about 50k CPU hours which already requires significant, cluster-sized computing infrastructure. On the other hand, some of the most demanding NNLO calculations performed to date \cite{Czakon:2021mjy,Alvarez:2023fhi} have required up to tens of millions of CPU hours, even with the relative maturity and multi-year optimization efforts they have benefited  from. For these reasons it is unlikely that fast, general, publicly available NNLO codes will appear in the near future.

With the clear demand for NNLO predictions, the question then arises how NNLO calculations can easily be made available to the general user. In this work we propose one possible solution, called \hightea.

What is \hightea? This is a web service which stores Monte Carlo events produced in the course of a NNLO calculation, such that they can be analyzed in new calculations with minimal effort. While the basic idea is not new \cite{Bern:2013zja,Andersen:2014qqa,Maitre:2018gua,Maitre:2020blv}, the utility of \hightea\ stems from the fact that it offers extensive functionality for efficiently storing, processing, analyzing and visualizing the results of such a calculation. 

Who is \hightea\ for? It is for {\it everyone}. \hightea\ is structured in such a way that it requires neither technical knowledge of higher-order perturbative calculations nor computing infrastructure or technical resources. 

In the following sections we explain in detail the features and workings of \hightea.

\section{\hightea\ in a nutshell}\label{sec:features}

\subsection{Utility of \hightea }

\hightea's utility can be summarized in the following way:

\begin{enumerate}
\item It allows the user to compute {\it any} infrared safe $n$-dimensional differential distribution in {\it any} process which has already been added to the library of available processes. The up-to-date list of all available processes can be obtained from \hightea\ itself (see sec.~\ref{sec:examples}).
\item The output of a \hightea\ computation is a histogram, and the input is the histogram's specification. In addition, \hightea\ automatically produces nice plots of the computed histograms. If experimental data is available, the user can directly be provided with a theory/data comparison plot, including all relevant theory uncertainties: MC, scale and pdf. This is in addition to the data uncertainty, if known. See sec.~\ref{sec:examples} for an example.
\item Predictions at present are in fixed-order perturbation theory, but can potentially be extended to include resummed/showered events.
\item \hightea\ is not tied to a specific perturbative order. Any perturbative order that can be computed with Monte Carlo methods can be included. At present, our main goal is to offer NNLO QCD accurate corrections for LHC processes. 
\item \hightea\ ``knows" most standard kinematic variables relevant for a given process. For added flexibility, users can additionally define their own kinematic variables in terms of commonly available mathematical functions \cite{Harris:2020xlr}.
\item \hightea\ offers the user {\it complete} control over the functional forms of the factorization and renormalization scales. This is achieved similarly to the user-defined kinematic variables discussed above. Scale variation is flexible, too; one can choose among the standard 3-point or 7-point variations or can define a custom variation.
\item Any pdf set available in the {\tt LHAPDF} library \cite{Buckley:2014ana} can be used. \hightea\ produces pdf variation based on either standard or specialized (recommended - see sec.~\ref{sec:examples}) pdf sets.
\item The jet size in processes involving jets can also be varied.  
\end{enumerate}

\subsection{\hightea\ is designed to be used by {\it everyone}}

Here are the main reasons why one would want to use \hightea\ :
\begin{enumerate}
\item With the help of \hightea\ one can perform their own state-of-the-art calculation without the need for complex computing codes or the knowledge of a technical subject like higher-order perturbative calculations.
\item It can be used to compute processes that may not be available in public codes or to cross-check and/or supplement existing codes and calculations.
\item No need for major computing infrastructures (typically, a large cluster). Any user can obtain their own state-of-the-art predictions that are equivalent to a full-fledged NNLO calculation using only their own personal computer or smart phone.
\item Predictions derived from \hightea\  are very fast. A typical calculation takes between seconds and an hour, depending on the setup.
\end{enumerate}

\subsection{Limitations of \hightea }

Like any other tool, \hightea\ has its limitations:
\begin{enumerate}
\item The values of parameters intrinsic to the collider or the process cannot be changed. Examples are the collider c.o.m. energy and the value of the top quark mass in processes involving top quarks. 
\item Predictions for processes with no intrinsic hard transverse scale (jet production, for example) are subject to a set of pre-defined generation-level cuts. These are needed for the definition of the process. The user does not need to be too concerned with those beyond ensuring that every event that is accepted by the user’s cuts is also accepted by the generation cuts.
\item Fixed Monte Carlo statistics. The statistics of theoretical events is fixed and the user cannot increase it. This means that if one requests, for example, very small bins and/or is interested in the tails of kinematic distributions, one may get predictions with unsuitably large MC uncertainty. \hightea\ always returns an estimate of the MC uncertainty, i.e. the user is always provided with a quantitative measure of how reliable the prediction is in terms of statistics. If needed, MC statistics can be increased by the authors of \hightea\ upon request. One should keep in mind that separate evaluations using the same dataset are not statistically independent and, for example, one cannot reduce the MC uncertainty by averaging multiple calculations.
\item Only processes that have been added to the library can be analyzed. If the process of interest is not available please request it from the \hightea\ authors. 
\end{enumerate}

The above limitations are not unusual and are, in fact, inherent to any Monte Carlo program. Here are some hints about how one can deal with some of these limitations. 

Some parameters like the value of the strong coupling constant, pdf and scales are at the complete disposal of the user. Other parameters, like the top quark mass, cannot be changed, yet predictions for a set of values are often required. A practical solution is to make available with \hightea\ three datasets that have different values of $m_t$: one around the world average and two more with $m_t$ values that are about $\pm1$ GeV away from it. The user can then interpolate on a per-bin basis between these 3 values to derive approximate prediction for any $m_t$ value within this range. An extensive application of this interpolation technique can be found in ref.~\cite{Cooper-Sarkar:2020twv}.

Whenever generation-level cuts are applied to a process, this is specified in the info file for that process (see sec.~\ref{sec:examples} for information about how to access it). As a rule, such cuts are applied to typical for that process variables like a suitable invariant mass or $H_T$. It is easy to ensure consistency when the user applies cuts to the same variable, because in such a case all one needs is a user-level cut that is more exclusive than the generation-level one. In case user-level cuts are applied to a different variable, the user must ensure consistency between the cuts. This can be a non-trivial task, however it is an important one for maintaining the correctness of the results.

\subsection{Under the hood: a gentle introduction to the inner workings of \hightea }

To be able to make state of the art calculations and take full advantage of \hightea\ one does not need to have any knowledge about how it works. The purpose of this section is to satisfy the curiosity of the reader about the inner workings of the program. 

\hightea\ analyzes precomputed and stored weighted ``theoretical events". An event is a set of parton id's, parton momenta and a weight. The weight represents the value of a probability density distribution in one point (specified by the set of momenta and partonic fractions $x_{1,2}$ and/or $z$). Beyond the leading order, an observable receives contributions from additional partonic channels and from final states with different multiplicities. Furthermore, the cancellation of infrared divergences necessitates the inclusion of so-called ``counterevents" (these are also weighted events but are derived from reduced matrix elements). For these reasons, in the context of MC event generation, a complete prediction is expressed not through a single event but by a set of events that typically have different dimensionalities. \hightea\ automatically accounts for all such contributions.

When a user submits a request for an analysis, \hightea\ analyzes all saved events and bins them accordingly. The user can modify parameters like the value of the strong coupling constant, the pdf set and the renormalization and factorization scales. This is achieved by re-weighting each event during the course of the analysis. In order to gain access to these scales, to $\alpha_S$ and to pdfs, an event's weight is saved as an array which contains the coefficients of all terms of the type $\log^n(\mu_F)\log^m(\mu_R)$, and factorizes the dependence on $\alpha_S$ and pdfs. It is then easy to reconstruct the complete weight for any value or functional form of these two scales, as well as for any value of the strong coupling constant and pdf set available via the {\tt LHAPDF} library.

We partially unweight events to reduce the size of the tables of events. We introduce a maximum weight to which we unweight using a hit-and-miss Monte-Carlo. Events with larger event weight than this maximum are kept using their weight. The value of the maximum weight is tuned to balance a high acceptance rate and a minimal number of weighted events. A higher acceptance rate translates directly to shorter computation times for a given sample size. At the same time, a smaller number of weighted events minimizes the resulting Monte Carlo uncertainty for a sample of a given size. The working point depends on the process and the optimization of the phase space sampling. Related ideas have been discussed in refs.~\cite{Nachman:2020fff,Matchev:2020jqz,Andersen:2023cku}.

Processes with jets deserve special attention. For such processes, \hightea\ retains the full parton-level information for all events/counterevents. Nonetheless, there is an implicit dependence on a jet algorithm and jet radius $R$ since, during the generation of the events, generation-level selection cuts are applied at the jet level. Once an event passes the jet-level selection requirement, all information about the event is saved at the parton-level. When the user requests a calculation with jets, the jet algorithm specified by the user is used to cluster the partons into jets. Correct results can only be guaranteed if the user uses the same jet algorithm as the algorithm used during the generation of the dataset. This information is available in the process info file.

\section{High-level overview of \hightea's structure and usage}\label{sec:high-level}

This section is the central part of this paper and should be read by all users. It gives a high-level overview of the structure and usage of this project and explains everything in non-technical, intuitive terms. The depth is sufficient so a user with basic knowledge can start using the project right away. 

\begin{figure}[t]
\begin{centering}
\includegraphics[width = 0.48\textwidth,trim=0 1mm 0 0]{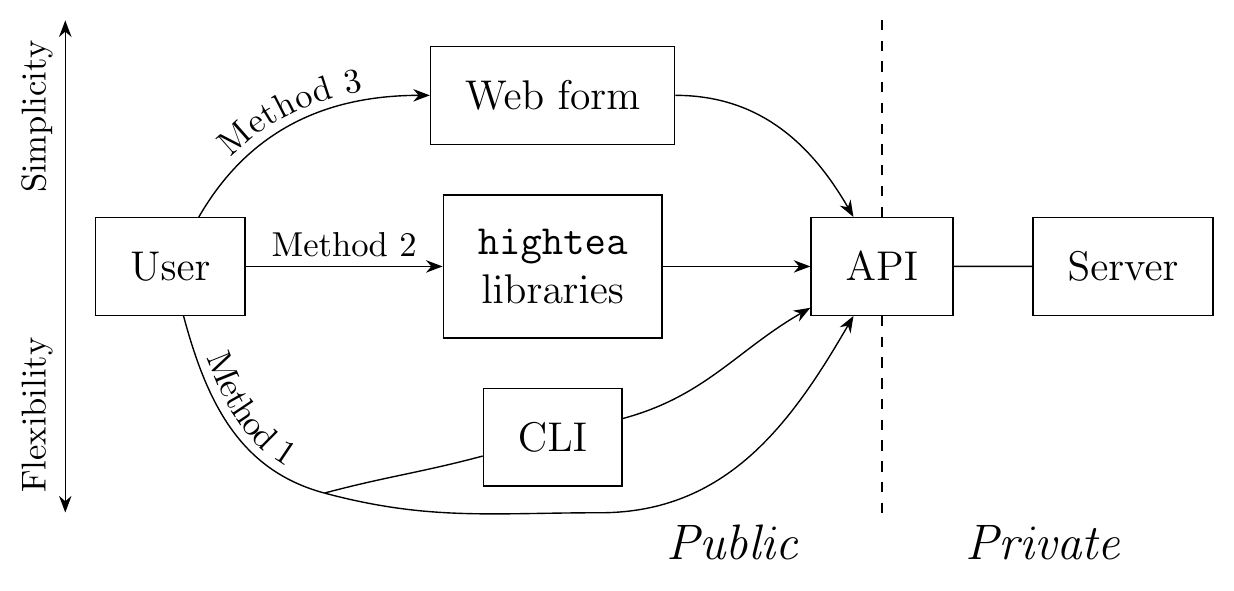}
\caption{High-level structure of \hightea. Shown are library's main components together with the three access methods. The vertical dashed line delineates the separation of public and private components.}
\label{fig:scheme}
\end{centering}
\end{figure}

From the point of view of the user, \hightea\ is naturally divided in two parts, see fig.~\ref{fig:scheme}:
\begin{itemize}
\item {\bf Server}: it is completely hidden from the user; the user has no direct access to it and in effect it should be thought of by the users as a black box,
\item {\bf Front end}: directly accessible by the user. It allows the user to submit requests and to receive and visualize results.
\end{itemize}

\subsection{Server}

We expect this section to be of limited interest and will provide a brief overview only. The server side consists of the databases for all precomputed processes as well as various software facilities that enable the analysis of these databases and the transfer of results and/or requests to/from the user interface. The sizes of the various databases vary greatly from process to process and can be between hundreds of GBytes to many TBytes. For ease of storage, analysis and server memory management, each dataset is split into a number of compressed files of approximately equal size. The database files are stored in the parquet format \cite{parquet}, which is designed for efficient data storage and retrieval.

A typical analysis can take between seconds and an hour, depending on the size of the database and how extensive the requested analysis is. The bulk of the time is spent not on reading the database but on the various mathematical operations needed to recompute scales and especially pdf sets. In fact, re-computations with different pdf sets tend to be the slowest part of any computation. The time is proportional to the number of pdf members used. For this reason we recommend the use of Specialized Minimal pdf (SMPDF) sets \cite{Carrazza:2016htc} whenever pdf variation is required and when such a set has been made available by the authors. 

The CPU and memory consumption for such an analysis can also be significant but are within the capabilities of a standard
\footnote{Currently, we are using a DELL PowerEdge R620 (2 x 8-core Xeon E5-2650v2) server with 64GB memory.}
server.

The software on the server side also deals with the queuing of requests and the storing of the final results. We have implemented a solution where all requests and results are stored long-term and can be retrieved by the user with the help of a token they are provided with when their request is submitted. For this reason, the user may disconnect once a job request is submitted and can retrieve the job at a later time.

\subsection{Front end}

The user interface offers three possible methods of accessing the database of \hightea, see fig.~\ref{fig:scheme}. These methods are designed with different types of users in mind:
\begin{enumerate}
\item Directly accessing the project's \API.
\item Accessing the project through a Python library. Suitable for the vast majority of users. It is expected to be the most utilized way for accessing the project.
\item Accessing the project via a web form. Suitable for every user. It is trivial to use however it offers less flexibility and sophistication and not all features can be utilized. This method is suitable for quick checks and demonstrations but less so for serious work.
\end{enumerate}

\subsubsection{\API}

The \API\ is the only point through which the server can be accessed. It utilizes standard web communication protocols and can be accesses from anywhere on the web via \JSON\ format queries. The complete description of the \API\ can be found in the automatically generated online document \cite{API}. Further details about the \API\ can be found here \cite{client-side}.

The \API\ allows the user to utilize every feature available in \hightea. \API's direct use, however, requires significant technical sophistication since the user needs to prepare their own query in a \JSON\ format, submit it to the corresponding web address, receive and store the token which uniquely identifies the request and, once the request has been processed, retrieve the results by querying the web access point and retrieving the results which are also in a \JSON\ format. 

\subsubsection{\tt CLI}

We have provided a Command Line Interface utility, described here \cite{client-side}, which simplifies the task of submitting \JSON\ requests to the \API.

\subsubsection{Web form}

For the vast majority of users we have designed two more access methods: via our web form and via our Python library.

The web form access \cite{webform} is self explanatory. It represents a standard web form where one can fill in the request, press a button, and then wait for the results to be displayed. Once the process is selected, on the right hand side of the page a detailed information about the process is displayed. In essence, this is the content of the info file mentioned above. Please note that the web form is under constant improvement and so it may evolve with time. Its current functionality is limited by web design constraints but we could imagine future improvements that will make it potentially as flexible and as useable as the Python library.

\subsubsection{Python library}\label{sec:python-library}

The most versatile and recommended way of accessing the project is via the Python library. A detailed description of the library can be found here \cite{Python-library}. The fact that the library is written in Python is not essential since, in order to use it, no knowledge of Python is required. We have created higher level wrapper functions which hide the Python language specifics and allow the user to interact with it using high-level intuitive commands. Of course, the user also has direct access to all low-level commands in the Python language. The user-friendliness and visual appeal is ensured by our embedding of the functionalities in a JupyterLab Notebook Interface \cite{Jupyter}. The way a user interacts with a Jupyter Notebook is very similar to the Notebooks of popular computer algebra systems like Mathematica \cite{Mathematica} and Maple \cite{Maple}. 

We have created two different ways of using this library. The first one involves its installation on a personal computer. This is a fairly standard process but it may require the installation of various standard Python libraries. This step should be fairly straightforward for people who are familiar with Python or the basics of installing software on Unix-like platforms. The second way is much easier since it involves no software installation at all. For this reason it is our default recommendation for using \hightea. In this second approach Jupyter Notebooks are run directly in the cloud. There are many websites and web-based services which allow users to run and collaborate on their own Jupyter Notebooks. One such option is the Google Research Colaboratory (a.k.a. Colab) \cite{Colab}. It offers the possibility to install, share and collaborate on Jupyter Notebooks. As of the writing of this paper this service is free and only requires a free Google account. On this page \cite{colab-access} we have provided a number of \hightea\ examples and a tutorial in the form of ready-to-use Jupyter Notebooks which the user can immediately run directly on Colab. The Notebooks and their results can be saved for later use on the user's Google Drive.

\section{Examples of uses and test cases}\label{sec:examples}

In this section we describe a typical workflow for using \hightea. It assumes the default Jupyter notebook approach for accessing \hightea\ described in sec.~\ref{sec:python-library}. The following discussion can be translated to the other ways of accessing \hightea\ in a rather straightforward way.

Before we delve into the specifics of how jobs are submitted and results retrieved, there is one general aspect of \hightea\ related to authentication that needs to be clarified. \hightea\ is meant to be used concurrently by a large number of users who may submit a large number of jobs nearly simultaneously and these jobs may take rather different times to execute. This aspect requires a system of labeling jobs and users over a potentially extended period of time. In \hightea\ we have implemented the following process. 

{\bf Job identification.}  When submitted, a job is assigned a unique ID in the form of a token. The token is returned to the user who submits the job and is displayed in the Jupyter notebook. They are stored in the {\tt .job} file associated with the given job. The token is needed since it allows users to exit their Notebook right after the job is submitted and then restart the notebook at a later time and retrieve the result. To achieve this the user simply executes all lines of the Notebook.  The Notebook will automatically submit the appropriate token to the \hightea\ server and retrieve the result. Results stored on the \hightea\ server currently do not expire.

{\bf User authentication.} Internally, \hightea\ keeps record of all submitted requests and outputs. To allow users to retrieve their past jobs even without having to keep track of all individual job tokens, we have introduced a user authentication system. If desired, a user can obtain a personalized token which needs to be used with all job submissions. This allows users to retrieve all their past jobs. An example can be found in the Notebook {\tt Example-authentication.ipynb} in \cite{colab-access}. At present users are not required to authenticate and can submit jobs as anonymous users. 

{\bf Notebook setup.} The example Notebooks available at \cite{colab-access} are ready to be run. Beforehand, the user needs to clone the desired Notebook to their own Google Drive, see \cite{Python-library} for details. A Notebook saved on the local Drive can directly be run, one does not need to go back to the page \cite{colab-access}. Users can create new Notebooks out of the several examples provided at \cite{colab-access} and use them in a way that suits their needs. 

{\bf The actual calculation.} In the rest of this section we introduce the \hightea\ functionality, using as an example the Notebook {\tt Example-ttbar.ipynb}. This Notebook consists of the following steps:
\begin{itemize}
\item Setup of the notebook,
\item Download experimental data from {\tt www.hepdata.net} for each one of the available differential distributions,
\item Convert the downloaded data to a format convenient for further processing,
\item Automatically extract from the data the binnings of each variable,
\item Specify the mathematical expressions for all measured observables,
\item For each order of interest (LO, NLO, NNLO, etc) prepare a job (called {\tt job} in this example) by giving it the following attributes, and then submit the job:
	\begin{itemize}
	\item Initialize {\tt job} by giving it a name,
	\item Specify the process ($t\bar t$ production at the LHC at 13 TeV in this case),
	\item Specify all one-dimensional distributions that will be computed,
	\item Specify if scale uncertainty is required (by saying what scale variation to use: 3-point, 7-point or custom),
	\item Submit {\tt job} to \hightea\ using the command {\tt job.request()},
	\item Place the results, when ready, in a new object called {\tt jobs}.
	\end{itemize}
\item Pre-format data and theory predictions for plotting (using a special Python function {\tt Run}; more info about it can be obtained by typing {\tt ?Run}),
\item Plot all observables.
\end{itemize}

In the plots, one can see data (as bars) together with LO, NLO and NNLO predictions. Both the absolute distributions and their ratio w/r to LO are plotted. Shown is the scale variation band as well as the MC uncertainty as a vertical bar for each curve and bin.

The above example does not exhaust all options and possibilities offered by \hightea. For example, the user can print the computed differential distributions in the form of tables using the command {\tt job.show\_result()} and then use them in their own plotting routines. Some of the other example Notebooks show its usage. 

Here are some more functionalities: 
\begin{itemize}
\item List all available processes: see the Notebook {\tt Example-lower-level-interface.ipynb},
\item Output the information about a process: {\tt job.process('pp\_tt\_13000\_172.5')} (for the specific process in the above example), 
\item Define new variables: {\tt job.define\_new\_variable('name','definition')}, 
\item The renormalization and factorization scales to be used: {\tt job.scales('$\mu_R$','$\mu_F$')}, 
\item Specify the pdf set to be used. For example: {\tt job.pdf('CT14nnlo')}). 
\item To introduce phase space cuts one uses, for example, {\tt job.cuts('pt\_t > 30')}, see the Tutorial Notebook for details.
\item Generic $n$-dimensional differential cross sections can be computed, see the example Notebook {\tt Example-ttbar-simple-2D.ipynb}.
\item For calculations of processes with jets see the Notebook {\tt Example-inclusive-jets.ipynb}.
\end{itemize}

One can also request pdf variation by typing {\tt job.pdf\_variation()}. The pdf prescription relevant for the requested pdf set is automatically used. A note about pdf variation and timings: LO and NLO calculations are very fast, and tend to be of the order of seconds. NNLO calculations are much slower and as a rule take about 10 to 100 times longer. As mentioned above, the speed of a calculation is directly related to the number of scales and pdf members used. For this reason when pdf variation is requested a calculation can be slowed down by a factor of about 100. We have implemented by default SMPDF sets, for the processes for which such sets are available, since such sets reduce the time for computing pdf variation by a factor of about 10. If full pdf variation is required one can specify this by typing {\tt job.pdf\_variation('full')}.

\section{Conclusions}

In this work we introduce \hightea: a novel approach for distributing and performing fast NNLO calculations for collider processes. The main idea is simple: the \hightea\ authors precompute ``theoretical" events using Monte Carlo methods, and store the events for later analysis. To avoid burdening users with technical aspects like transferring, storing and analyzing vast amounts of data, we have developed the complete software and hardware infrastructure required for dealing with these tasks. The only thing the user needs to do is to request a calculation via the web and then receive the result in the form of a binned distribution, accompanied by a nice plot. This way, the user is completely shielded from the intricacies of doing NNLO calculations, the need for securing huge computing resources and from having to have even basic programming experience. In fact, as of the writing of this article, a user only needs a free Google account and a device for accessing the Internet (like a personal computer or a smart phone) in order to request and receive, {\it typically within minutes}, a most sophisticated calculation. For further details, and to access ready-to-run examples, please visit \hightea's homepage:
\begin{center}
\url{https://www.precision.hep.phy.cam.ac.uk/hightea}
\end{center}

The main restrictions when using \hightea\ are that one can only compute processes that have been already been implemented by \hightea's authors, and that the statistics of events is fixed, i.e. the user cannot increase it. Our hope is to keep adding new processes on a regular basis. The number of events included in the various datasets is chosen in such a way that at NNLO the MC error is subdominant to the scale one for a typical experimental LHC setup. The number of events can be increased upon request. 

All processes which have been included in \hightea\ have been carefully validated versus independent calculations available in the literature. The relevant papers are provided with the info files accompanying each dataset. When using \hightea\ please cite the present publication together with the said papers.

\hightea\ offers new opportunities in terms of access to cutting precision perturbative calculations for collider processes. For example, the lack until now of accessible NNLO predictions has often precluded the usage of such calculations in various searches and related LHC analyses. \hightea\ is meant to effectively close this gap. \hightea\ can handle distributions of any dimension and any infrared safe observable, and its predictions are very fast. 

\hightea\ offers new possibilities for comparing theory to data which is available in alternative and more flexible formats like the unbinned distributions recently proposed in ref.~\cite{Arratia:2021otl}, or the CMS Open Data initiative \cite{CMS-opendata}. \hightea\ makes it particularly easy to automatically compare theoretical predictions to data which is available in the HEPData repository \cite{hepdata}. A working example is provided.

There are many more potential uses of \hightea, all of which are reasonably easy to implement and pursue: 
\begin{itemize}
\item Allow other providers to add their own calculations to the \hightea. Examples of potential uses are NLO calculations of EW corrections and SMEFT, like refs.~\cite{Giani:2023gfq,Cepedello:2023yao,Moretti:2023dlx,Kassabov:2023hbm}. 
\item Allow users to access individual pdf channels; incorporate scale variations correlated with pdfs~\cite{Kassabov:2022orn}.
\item Allow users to access contributions from specific powers of couplings (i.e. not just LO, NLO, etc). Such a feature would be relevant for calculations in the full Standard Model or beyond. 
\item \hightea\ is very well suited for adding contributions from EFT operators in addition to the usual QCD or SM contributions. This will allow users to directly request an analysis in a complete SMEFT framework.
\item In addition to bins and plots, it is relatively easy to modify \hightea\ by extending our implementation \cite{Czakon:2017dip} so it can also output {\tt fastNLO} \cite{fastnlo,Britzger:2012bs} or related \cite{Carli:2010rw,Carrazza:2020gss,Buckley:2020bxg} grids.
\end{itemize}

Finally, we would like to stress that while \hightea\ has been developed in the LHC era and, naturally, is currently focused on predictions for LHC physics, it can equally well handle other colliders. Given the increasing interest in future $e^+e^-$ colliders, \hightea\ applications to $e^+e^-$ processes will be particularly useful.

\begin{acknowledgments}
We thank Steve Wotton for his server infrastructure support. The work of M.C. was supported by the Deutsche Forschungsgemeinschaft under grant 396021762 - TRR 257; support by a grant of the Bundesministerium für Bildung und Forschung (BMBF) is additionally acknowledged. The research of Z.K., A.M., R.P. and A.P. has received funding from the European Research Council (ERC) under the European Union's Horizon 2020 Research and Innovation Programme (grant agreement no. 683211). Z.K. is also supported by the European Research Council under the European Union’s Horizon 2020 research and innovation Programme (grant agreement n.950246). A.M. was also supported by the UK STFC grants ST/L002760/1 and ST/K004883/1. R.P. acknowledges support from the Leverhulme Trust and the Isaac Newton Trust. A.M. and R.P. acknowledge the use of the DiRAC Cumulus HPC facility under Grant No. PPSP226.
\end{acknowledgments}


\begin{thebibliography}{99}

\bibitem{Frixione:1995ms}
S.~Frixione, Z.~Kunszt and A.~Signer,
Nucl. Phys. B \textbf{467}, 399-442 (1996)
[arXiv:hep-ph/9512328 [hep-ph]].

\bibitem{Catani:1996vz}
S.~Catani and M.~H.~Seymour,
Nucl. Phys. B \textbf{485}, 291-419 (1997)
[erratum: Nucl. Phys. B \textbf{510}, 503-504 (1998)]
[arXiv:hep-ph/9605323 [hep-ph]].

\bibitem{Nagy:2003tz}
Z.~Nagy,
Phys. Rev. D \textbf{68}, 094002 (2003)
[arXiv:hep-ph/0307268 [hep-ph]].

\bibitem{Campbell:2011bn}
J.~M.~Campbell, R.~K.~Ellis and C.~Williams,
JHEP \textbf{07}, 018 (2011)
[arXiv:1105.0020 [hep-ph]].

\bibitem{Campbell:2015qma}
J.~M.~Campbell, R.~K.~Ellis and W.~T.~Giele,
Eur. Phys. J. C \textbf{75}, no.6, 246 (2015)
[arXiv:1503.06182 [physics.comp-ph]].

\bibitem{Campbell:2019dru}
J.~Campbell and T.~Neumann,
JHEP \textbf{12}, 034 (2019)
[arXiv:1909.09117 [hep-ph]].

\bibitem{Alwall:2014hca}
J.~Alwall, R.~Frederix, S.~Frixione, V.~Hirschi, F.~Maltoni, O.~Mattelaer, H.~S.~Shao, T.~Stelzer, P.~Torrielli and M.~Zaro,
JHEP \textbf{07}, 079 (2014)
[arXiv:1405.0301 [hep-ph]].

\bibitem{Gleisberg:2008ta}
T.~Gleisberg, S.~Hoeche, F.~Krauss, M.~Schonherr, S.~Schumann, F.~Siegert and J.~Winter,
JHEP \textbf{02}, 007 (2009)
[arXiv:0811.4622 [hep-ph]].

\bibitem{Sherpa:2019gpd}
E.~Bothmann \textit{et al.} [Sherpa],
SciPost Phys. \textbf{7}, no.3, 034 (2019)
[arXiv:1905.09127 [hep-ph]].

\bibitem{Frixione:2007vw}
S.~Frixione, P.~Nason and C.~Oleari,
JHEP \textbf{11}, 070 (2007)
[arXiv:0709.2092 [hep-ph]].

\bibitem{Sjostrand:2014zea}
T.~Sj\"ostrand, S.~Ask, J.~R.~Christiansen, R.~Corke, N.~Desai, P.~Ilten, S.~Mrenna, S.~Prestel, C.~O.~Rasmussen and P.~Z.~Skands,
Comput. Phys. Commun. \textbf{191}, 159-177 (2015)
[arXiv:1410.3012 [hep-ph]].

\bibitem{Anastasiou:2015vya}
C.~Anastasiou, C.~Duhr, F.~Dulat, F.~Herzog and B.~Mistlberger,
Phys. Rev. Lett. \textbf{114}, 212001 (2015)
[arXiv:1503.06056 [hep-ph]].

\bibitem{Chen:2021isd}
X.~Chen, T.~Gehrmann, E.~W.~N.~Glover, A.~Huss, B.~Mistlberger and A.~Pelloni,
Phys. Rev. Lett. \textbf{127}, no.7, 072002 (2021)
[arXiv:2102.07607 [hep-ph]].

\bibitem{Camarda:2021ict}
S.~Camarda, L.~Cieri and G.~Ferrera,
Phys. Rev. D \textbf{104}, no.11, L111503 (2021)
[arXiv:2103.04974 [hep-ph]].

\bibitem{Chen:2021vtu}
X.~Chen, T.~Gehrmann, N.~Glover, A.~Huss, T.~Z.~Yang and H.~X.~Zhu,
Phys. Rev. Lett. \textbf{128}, no.5, 052001 (2022)
[arXiv:2107.09085 [hep-ph]].

\bibitem{Chen:2022cgv}
X.~Chen, T.~Gehrmann, E.~W.~N.~Glover, A.~Huss, P.~F.~Monni, E.~Re, L.~Rottoli and P.~Torrielli,
Phys. Rev. Lett. \textbf{128}, no.25, 252001 (2022)
[arXiv:2203.01565 [hep-ph]].

\bibitem{Neumann:2022lft}
T.~Neumann and J.~Campbell,
Phys. Rev. D \textbf{107}, no.1, L011506 (2023)
[arXiv:2207.07056 [hep-ph]].

\bibitem{Duhr:2020sdp}
C.~Duhr, F.~Dulat and B.~Mistlberger,
JHEP \textbf{11}, 143 (2020)
[arXiv:2007.13313 [hep-ph]].

\bibitem{Baglio:2022wzu}
J.~Baglio, C.~Duhr, B.~Mistlberger and R.~Szafron,
JHEP \textbf{12}, 066 (2022)
[arXiv:2209.06138 [hep-ph]].

\bibitem{Camarda:2023dqn}
S.~Camarda, L.~Cieri and G.~Ferrera,
[arXiv:2303.12781 [hep-ph]].

\bibitem{Gehrmann-DeRidder:2005btv}
A.~Gehrmann-De Ridder, T.~Gehrmann and E.~W.~N.~Glover,
JHEP \textbf{09}, 056 (2005)
[arXiv:hep-ph/0505111 [hep-ph]].

\bibitem{Currie:2016bfm}
J.~Currie, E.~W.~N.~Glover and J.~Pires,
Phys. Rev. Lett. \textbf{118}, no.7, 072002 (2017)
[arXiv:1611.01460 [hep-ph]].

\bibitem{Gaunt:2015pea}
J.~Gaunt, M.~Stahlhofen, F.~J.~Tackmann and J.~R.~Walsh,
JHEP \textbf{09}, 058 (2015)
[arXiv:1505.04794 [hep-ph]].

\bibitem{DelDuca:2016csb}
V.~Del Duca, C.~Duhr, A.~Kardos, G.~Somogyi and Z.~Tr\'ocs\'anyi,
Phys. Rev. Lett. \textbf{117}, no.15, 152004 (2016)
[arXiv:1603.08927 [hep-ph]].

\bibitem{Catani:2007vq}
S.~Catani and M.~Grazzini,
Phys. Rev. Lett. \textbf{98}, 222002 (2007)
[arXiv:hep-ph/0703012 [hep-ph]].

\bibitem{Boughezal:2015dva}
R.~Boughezal, C.~Focke, X.~Liu and F.~Petriello,
Phys. Rev. Lett. \textbf{115}, no.6, 062002 (2015)
[arXiv:1504.02131 [hep-ph]].

\bibitem{Czakon:2010td}
M.~Czakon,
Phys. Lett. B \textbf{693}, 259-268 (2010)
[arXiv:1005.0274 [hep-ph]].

\bibitem{Czakon:2014oma}
M.~Czakon and D.~Heymes,
Nucl. Phys. B \textbf{890}, 152-227 (2014)
[arXiv:1408.2500 [hep-ph]].

\bibitem{Caola:2017dug}
F.~Caola, K.~Melnikov and R.~R\"ontsch,
Eur. Phys. J. C \textbf{77}, no.4, 248 (2017)
[arXiv:1702.01352 [hep-ph]].

\bibitem{Magnea:2018hab}
L.~Magnea, E.~Maina, G.~Pelliccioli, C.~Signorile-Signorile, P.~Torrielli and S.~Uccirati,
JHEP \textbf{12}, 107 (2018)
[erratum: JHEP \textbf{06}, 013 (2019)]
[arXiv:1806.09570 [hep-ph]].

\bibitem{Boughezal:2016wmq}
R.~Boughezal, J.~M.~Campbell, R.~K.~Ellis, C.~Focke, W.~Giele, X.~Liu, F.~Petriello and C.~Williams,
Eur. Phys. J. C \textbf{77}, no.1, 7 (2017)
[arXiv:1605.08011 [hep-ph]].

\bibitem{Grazzini:2017mhc}
M.~Grazzini, S.~Kallweit and M.~Wiesemann,
Eur. Phys. J. C \textbf{78}, no.7, 537 (2018)
[arXiv:1711.06631 [hep-ph]].

\bibitem{Czakon:2021mjy}
M.~Czakon, A.~Mitov and R.~Poncelet,
Phys. Rev. Lett. \textbf{127}, no.15, 152001 (2021)
[erratum: Phys. Rev. Lett. \textbf{129}, no.11, 119901 (2022); erratum: Phys. Rev. Lett. \textbf{129}, no.11, 119901 (2022)]
[arXiv:2106.05331 [hep-ph]].

\bibitem{Alvarez:2023fhi}
M.~Alvarez, J.~Cantero, M.~Czakon, J.~Llorente, A.~Mitov and R.~Poncelet,
JHEP \textbf{03}, 129 (2023)
[arXiv:2301.01086 [hep-ph]].

\bibitem{Bern:2013zja}
Z.~Bern, L.~J.~Dixon, F.~Febres Cordero, S.~H\"oche, H.~Ita, D.~A.~Kosower and D.~Maitre,
Comput. Phys. Commun. \textbf{185}, 1443-1460 (2014)
[arXiv:1310.7439 [hep-ph]].

\bibitem{Andersen:2014qqa}
J.~R.~Andersen, S.~Bartle, Z.~Bern, F.~Febres Cordero, S.~H\"oche, D.~A.~Kosower, H.~Ita, N.~A.~Lo Presti, D.~Ma\^\i{}tre and K.~Ozeren,
PoS \textbf{LL2014}, 079 (2014)
[arXiv:1407.1621 [hep-ph]].

\bibitem{Maitre:2018gua}
D.~Ma\^\i{}tre,
J. Phys. Conf. Ser. \textbf{1085}, no.5, 052017 (2018)

\bibitem{Maitre:2020blv}
D.~Ma\^\i{}tre,
J. Phys. Conf. Ser. \textbf{1525}, no.1, 012014 (2020)

\bibitem{Harris:2020xlr}
C.~R.~Harris, K.~J.~Millman, S.~J.~van der Walt, R.~Gommers, P.~Virtanen, D.~Cournapeau, E.~Wieser, J.~Taylor, S.~Berg and N.~J.~Smith, \textit{et al.}
Nature \textbf{585}, no.7825, 357-362 (2020)
[arXiv:2006.10256 [cs.MS]].

\bibitem{Buckley:2014ana}
A.~Buckley, J.~Ferrando, S.~Lloyd, K.~Nordstr\"om, B.~Page, M.~R\"ufenacht, M.~Sch\"onherr and G.~Watt,
Eur. Phys. J. C \textbf{75}, 132 (2015)
[arXiv:1412.7420 [hep-ph]].

\bibitem{Cooper-Sarkar:2020twv}
A.~M.~Cooper-Sarkar, M.~Czakon, M.~A.~Lim, A.~Mitov and A.~S.~Papanastasiou,
[arXiv:2010.04171 [hep-ph]].

\bibitem{Andersen:2023cku}
J.~R.~Andersen, A.~Maier and D.~Ma\^\i{}tre,
[arXiv:2303.15246 [hep-ph]].

\bibitem{Nachman:2020fff}
B.~Nachman and J.~Thaler,
Phys. Rev. D \textbf{102}, no.7, 076004 (2020)
[arXiv:2007.11586 [hep-ph]].

\bibitem{Matchev:2020jqz}
K.~T.~Matchev and P.~Shyamsundar,
SciPost Phys. \textbf{10}, no.2, 034 (2021)
[arXiv:2006.16972 [hep-ph]].

\bibitem{parquet}
Apache Parquet data file format \url{https://parquet.apache.org}

\bibitem{Carrazza:2016htc}
S.~Carrazza, S.~Forte, Z.~Kassabov and J.~Rojo,
Eur. Phys. J. C \textbf{76}, no.4, 205 (2016)
[arXiv:1602.00005 [hep-ph]].

\bibitem{API}
Online documentation of the \hightea\ \API\ \url{https://www.hep.phy.cam.ac.uk/hightea/docs}

\bibitem{client-side}
Complete online \hightea\ client-side documentation \url{https://hightea-client.readthedocs.io}

\bibitem{webform}
Access to \hightea\ via webform \url{https://www.hep.phy.cam.ac.uk/hightea/webform/}

\bibitem{Python-library}
Access to \hightea\ via our Python library \url{https://www.precision.hep.phy.cam.ac.uk/hightea/google-colab/}

\bibitem{Jupyter}
The open-source software project Jupyter \url{https://jupyter.org}

\bibitem{Mathematica}
Wolfram Mathematica \url{https://www.wolfram.com/mathematica/}

\bibitem{Maple}
Maplesoft Maple \url{https://www.maplesoft.com/products/Maple/}

\bibitem{Colab}
Google Research Colaboratory \url{https://colab.research.google.com}

\bibitem{colab-access}
Access \hightea\ at Google Colab here: \url{https://colab.research.google.com/github/HighteaCollaboration/hightea-examples/blob/master/Start.ipynb}

\bibitem{Arratia:2021otl}
M.~Arratia, A.~Butter, M.~Campanelli, V.~Croft, D.~Gillberg, A.~Ghosh, K.~Lohwasser, B.~Malaescu, V.~Mikuni and B.~Nachman, \textit{et al.}
JINST \textbf{17}, no.01, P01024 (2022)
[arXiv:2109.13243 [hep-ph]].

\bibitem{CMS-opendata}
CMS Open Data initiative \url{http://opendata.cern.ch}

\bibitem{hepdata}
HEPData repository of collider data \url{https://www.hepdata.net}

\bibitem{Cepedello:2023yao}
R.~Cepedello, F.~Esser, M.~Hirsch and V.~Sanz,
[arXiv:2302.03485 [hep-ph]].
	
\bibitem{Giani:2023gfq}
T.~Giani, G.~Magni and J.~Rojo,
[arXiv:2302.06660 [hep-ph]].

\bibitem{Moretti:2023dlx}
S.~Moretti, L.~Panizzi, J.~Sj\"olin and H.~Waltari,
[arXiv:2302.03401 [hep-ph]].

\bibitem{Kassabov:2023hbm}
Z.~Kassabov, M.~Madigan, L.~Mantani, J.~Moore, M.~M.~Alvarado, J.~Rojo and M.~Ubiali,
[arXiv:2303.06159 [hep-ph]].

\bibitem{Kassabov:2022orn}
Z.~Kassabov, M.~Ubiali and C.~Voisey,
JHEP \textbf{03}, 148 (2023)
[arXiv:2207.07616 [hep-ph]].

\bibitem{Czakon:2017dip}
M.~Czakon, D.~Heymes and A.~Mitov,
[arXiv:1704.08551 [hep-ph]].

\bibitem{Britzger:2012bs}
D.~Britzger \textit{et al.} [fastNLO],
[arXiv:1208.3641 [hep-ph]].

\bibitem{fastnlo}
The {\tt fastNLO} interpolation grids library \url{https://fastnlo.hepforge.org}

\bibitem{Carli:2010rw}
T.~Carli, D.~Clements, A.~Cooper-Sarkar, C.~Gwenlan, G.~P.~Salam, F.~Siegert, P.~Starovoitov and M.~Sutton,
Eur. Phys. J. C \textbf{66}, 503-524 (2010)
[arXiv:0911.2985 [hep-ph]].

\bibitem{Carrazza:2020gss}
S.~Carrazza, E.~R.~Nocera, C.~Schwan and M.~Zaro,
JHEP \textbf{12}, 108 (2020)
[arXiv:2008.12789 [hep-ph]].

\bibitem{Buckley:2020bxg}
A.~Buckley, A.~Kvellestad, A.~Raklev, P.~Scott, J.~V.~Sparre, J.~Van Den Abeele and I.~A.~Vazquez-Holm,
Eur. Phys. J. C \textbf{80}, no.12, 1106 (2020)
[arXiv:2006.16273 [hep-ph]].

\end{thebibliography}
\end{document}